\documentclass[a4paper,fleqn]{cas-sc}

\usepackage{graphicx}

\newcommand{\dd}{\,\mathrm{d}}

\usepackage[numbers]{natbib}

\def\tsc#1{\csdef{#1}{\textsc{\lowercase{#1}}\xspace}}
\tsc{WGM}
\tsc{QE}
\tsc{EP}
\tsc{PMS}
\tsc{BEC}
\tsc{DE}
\begin{document}
\let\WriteBookmarks\relax
\def\floatpagepagefraction{1}
\def\textpagefraction{.001}
\shorttitle{Slow dynamics and ergodicity}
\shortauthors{LF Souza and TM Rocha Filho}

%\title{Slow dynamics and ergodicity in the one-dimensional self-gravitating system}
\title [mode = title]{Slow dynamics and ergodicity in the one-dimensional self-gravitating system}

\author[1]{LF Souza}[auid=001,bioid=2]

\author[2]{TM Rocha Filho}[auid=000,bioid=1,
                        orcid=0000-0002-7865-4085]

\address[1]{Centro de Ci\^encias Exatas e das Tecnologias, Universidade Federal do Oeste da Bahia and Instituto de F\'isica, Universidade de Bras\'ilia,
UnB - Brasília, DF, 70297-400}

\address[2]{Instituto de F\'\i sica and International Center for Condensed Matter Physics\\ Universidade de
Bras\'\i{}lia, CP: 04455, 70919-970 - Bras\'\i{}lia, Brazil}

%\author{L.~F.~Souza}
%\address{Centro de Ci\^encias Exatas e das Tecnologias, Universidade Federal do Oeste da Bahia and Instituto de F\'isica, Universidade de Bras\'ilia,
%UnB - Brasília, DF, 70297-400}
%\author{T.~M.~Rocha Filho}
%\address{Instituto de F\'\i sica and International Center for Condensed Matter Physics\\ Universidade de
%Bras\'\i{}lia, CP: 04455, 70919-970 - Bras\'\i{}lia, Brazil}

%\date{}

\begin{abstract}
We revisit the dynamics of the one-dimensional self-gravitating sheets models. We show that homogeneous and
non-homogeneous states have different ergodic properties. The former is non-ergodic and the one-particle distribution function has
a zero collision term if a proper limit is taken for the periodic boundary conditions.
Non-homogeneous states are ergodic in a time window of the order of the relaxation time to equilibrium, as similarly observe in
other systems with a long range interaction. For the sheets model this relaxation time
is much larger than other systems with long range interactions if compared to the initial violent relaxation time.
\end{abstract}

%\pacs{05.20.Dd; 05.70.Ln; 05.70.Ln}

\maketitle

\section{Introduction}
\label{sec:level1}

Lower dimensional models retaining the main characteristics of realistic systems has always been an important tool to grasp the phenomenology in
Statistical Physics.
They have been particularly important in understanding the non-equilibrium dynamics and equilibrium properties of systems with long range interactions,
which often present unusual properties not observed if the interaction is short-ranged, as
non-ergodicity, anomalous diffusion, non-Gaussian quasi-stationary states, negative microcanonical
heat capacity, ensemble inequivalence, and a very long relaxation time to thermodynamic equilibrium,
diverging with the particle number $N$~\cite{newbook,proc1,proc2,proc3,phyr,entropia,steiner,
scaling,lourenco,ergo1,ergo3,nv1,nv2,nv3,jain,transitions1,transitions2}.
Some one-dimensional models have been extensively studied in the literature, such as one-dimensional plasmas~\cite{dawson},
one-dimensional self-gravitating systems: the sheets and shell models~\cite{miller}, and derived models,
e.~g.\ the Ring~\cite{sota} and the Hamiltonian Mean Field (HMF) models~\cite{hmforig}. The dynamics of
systems with long range interactions can typically be divided in three stages: a violent collisionless relaxation from the initial condition into
a quasi-stationary state (or an oscillating state close to it), occurring in a very short time~\cite{lyndenbell},
followed by a very slow evolution to thermodynamic equilibrium, caused by the small cumulative effects of collisions (graininess).
The final and third stage is the thermodynamic equilibrium, that may never be attained in the
$N\rightarrow\infty$ limit, when the mean-field description becomes exact and the collisional contributions to the Kinetic equation vanish.
In this limit, and under suitable conditions, the dynamics is exactly described by the Vlasov equation~\cite{braun,steiner}.

Let us consider a system of identical particles described by the Hamiltonian:
\begin{equation}
H=\sum_{i=1}^N\frac{{\bf p}_i^2}{2m}+\frac{1}{N}\sum_{i<j=1}^N V_{ij},
\label{genlongham}
\end{equation}
with the interparticle potential $V_{ij}\equiv V(|\mathbf{r}_{i}-\mathbf{r}_{j}|)$, ${\bf p}_i$, ${\bf r}_i$
the momentum and positions for particle $i$, respectively, and
$m$ the mass of the particles. The factor $1/N$ in the potential energy term in Eq.~(\ref{genlongham})
is introduced such that the total energy is extensive~\cite{kac} (the so-called Kac factor). The one-particle distribution function $f({\bf p}, {\bf r};t)$
then satisfies the Vlasov equation:
\begin{equation}
\dot f\equiv\frac{df}{dt}=\frac{\partial f}{\partial t}+\frac{\bf p}{m}\cdot\frac{\partial f}{\partial\bf r}+{\bf F}\cdot\frac{\partial f}{\partial\bf p}=0,
\label{vlasoveq}
\end{equation}
where the mean-field force is given by:
\begin{equation}
{\bf F}({\bf r};t)=-\frac{\partial}{\partial\bf r}\int V({\bf r}-{\bf r}^\prime) f({\bf r}^\prime,{\bf p}^\prime;t)
\:\dd{\bf p}^\prime\:\dd{\bf r}^\prime.
\label{meanfieldfv}
\end{equation}
Collisional effects modify the Vlasov equation such that $\dot f=I[f]$ where the collisional integral $I[f]$ is a functional of $f$, usually obtained
using some approximation such as the weak coupling limit, the interparticle force is taken to be of order $\alpha\ll1$ and $I[f]$ is computed up to order $\alpha^2$,
or retaining terms of order $1/N$. The resulting kinetic equations are
called respectively the Landau and Balescu-Lenard equations~\cite{balescu}. For one-dimensional systems the collisional integral
in the Balescu-Lenard, Landau and Boltzmann equations vanish identically
in a homogeneous state and one must go to the next term in the approximation, i.~e.\ by computing $I[f]$
up to order $\alpha^3$ or $1/N^2$~\cite{sano,fouvry,scaling,lourenco}.

Let us consider the example of a system with a vanishing collisional integral for both homogeneous and non-homogeneous states
is given by $N$ identical particles in one dimension interacting only through zero-distance hard-core potential.
In this case the interaction causes a swap of particle velocities, and by simply relabeling the particles at
the moment of the collision one obtains a statistically equivalent system of free particles such that the one-particle distribution
function only evolves due to the free flux, and the corresponding kinetic equation if then given by
the one-dimensional Liouville equation with zero force:
\begin{equation}
        \frac{\partial f}{\partial t}+\frac{p}{m}\:\frac{\partial f}{\partial x}=0,
        \label{1dhard}
\end{equation}
where $m$ is the mass, $x$ is the position, and $p$ the momentum.
For a homogeneous state the one-particle distribution function is strictly constant, i.~e.\
the collisional integral vanishes identically.

Another yet simple model, but with long range interacting particles and real collisions (due to the discontinuity in the force at zero distance) is the
one-dimensional self-gravitating system of identical particle with unit mass and Hamiltonian~\cite{yawn}
\begin{equation}
	H=\sum_{i=1}^N\frac{p_i^2}{2}+\frac{1}{N}\sum_{i<j=1}^N \left|x_i-x_j\right|.
\label{ham1dgrav}
\end{equation}
The force on particle $i$ is given by $F_i=(N_-^{(i)}-N_+^{(i)})/N$, where $N_+^{(i)}$ and $N_-^{(i)}$ are the number of particles
to the right and left of the particle $i$, respectively, and particles can cross each other freely. The potential in this Hamiltonian
is obtained from the solution of the Poisson equation in one spatial dimension and corresponds to a system of $N$ infinite sheets with total finite mass.
The dynamics of this model has been studied in the literature in the last few decades, with the recurrent question if the system does relax to thermodynamic
equilibrium, due to the extreme slow slow dynamics of its macroscopic parameters~\cite{reidl,gouda,yawn,valageas,gabrieli}.
Joyce and Worrakitpoonpon introduced an order parameter to measure the distance to equilibrium and showed that this system in a non-homogeneous state
evolves to thermodynamic equilibrium~\cite{worra}. They showed this for a number of particles up to $N=800$, and yet requiring a very large simulation time to observe
the complete relaxation. This implies that the contribution of the collisional integral of the corresponding kinetic equation is very small.

The very slow relaxation towards equilibrium also manifests in the ergodic properties of the system.
%in the sense that a system is ergodic only
%if one considers a time window of the order of the relaxation time to equilibrium, which is usulally very large for systems with long range
%interactions.
A system with long range interactions is ergodic if averages of observables
over the history of a single particle are equal to the ensemble average, i.~e. to an average computed at a fixed time for the $N$ particles in the system.
This approach was used for the HMF model~\cite{ergo1,ergo2} and for a two-dimensional self-gravitating system~\cite{ergo3}.
In the limit $N\rightarrow\infty$ these systems are non-ergodic, and never reach the true thermodynamic equilibrium,
while for finite $N$ they are ergodic only after a time window of the order of the relaxation time to equilibrium.
Here we show that this results are also valid for the one-dimensional self-gravitating system with Hamiltonian in Eq.~(\ref{ham1dgrav}) in
a non-homogeneous state, but not the homogeneous case.
Indeed in the former, we show that by properly considering periodic boundary conditions and then taking the limit of the size of the unit cell
going to infinity, while keeping the density constant, the one-particle distribution function does not evolve in time, i.~e.\ the collisional
effects vanish.

The paper is structured as follows: in Section~\ref{sec2} we discuss separately the ergodic properties of
homogeneous and non-homogeneous states of the sheets model. The kinetic equation for the
homogeneous state is obtained in Sec.~\ref{sec3} with identically vanishing collisional contributions.
We close the paper with some concluding remarks in Sec.~\ref{sec4}.

\section{Slow dynamics and ergodicity}
\label{sec2}

We investigate the ergodic properties of the sheets model system using the approach in Ref.~\cite{ergo3}.
The system is ergodic if time averages taken over a given time window of length $t_e$ equals the ensemble average over the $N$-particles at
this same fixed time $t_e$, which we call ergodicity time.
We define the time average of the momentum of the $k$-th particle:
\begin{equation}
	\overline{p}_k(t)=\frac{1}{M}\sum_{j=1}^Mp_k(j\Delta t),
	\label{avgpdef}
\end{equation}
and similarly the time average of its position:
\begin{equation}
        \overline{x}_k(t)=\frac{1}{M}\sum_{j=1}^Mx_k(j\Delta t),
        \label{avgxdef}
\end{equation}
with a fixed time step $\Delta t$, $M=t/\Delta t$. We also consider the time dependent standard deviations
(supposing the averages over all particles vanish $\langle x\rangle=0$ and $\langle p\rangle=0$):
\begin{equation}
	\sigma_p\equiv\sqrt{\frac{1}{N}\sum_{k=1}^N\overline{p}_k^2(t)},
	\label{sigp}
\end{equation}
and
\begin{equation}
        \sigma_x\equiv\sqrt{\frac{1}{N}\sum_{k=1}^N\overline{x}_k^2(t)}.
        \label{sigx}
\end{equation}
Ergodicity for a system with long range interaction is then equivalent to~\cite{ergo3}
\begin{equation}
	\sigma_p(t)\rightarrow0\,\ \textrm{and}\,\, \sigma_x(t)\rightarrow0\,\,\textrm{for}\,\, t\rightarrow t_e.
	\label{defergosig}
\end{equation}
It was shown for the HMF model and for a two-dimensional self-gravitating system that $t_e\approx t_r$, with $t_r$ the relaxation time to
thermodynamic equilibrium~\cite{ergo3,ergo1,ergo2}.

We now consider separately the ergodic properties of non-homogeneous and homogeneous states of the sheets model.

\subsection{Non-homogeneous state}

\begin{figure}[ht]
	\begin{center}
	\scalebox{0.3}{\includegraphics{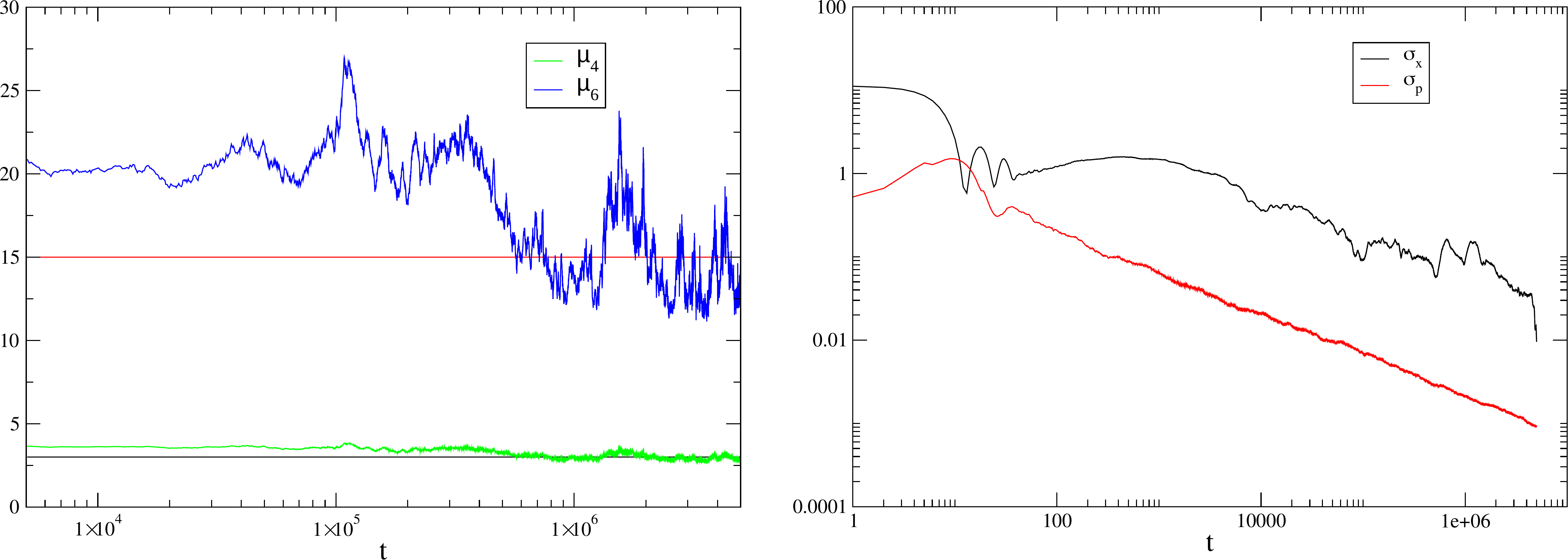}}
	\end{center}
%        \scalebox{0.3}{\includegraphics{moments_N100.pdf}}
%	\scalebox{0.3}{\includegraphics{sigma.pdf}}
	\caption{Left: Reduced moments $\mu_4$ and $\mu_6$ as a function of time for $N=100$ and a waterbag initial state with
	$x_0=10.0$ and $p_0=0.5$ for the system with Hamiltonian in Eq.~(\ref{ham1dgrav}) and open boundary conditions.
	A running average was performed over a time window of $\delta t=10\:000$. The straight lines
	correspond to the equilibrium values of $\mu_4=3$ and $\mu_6=15$ introduced for comparison purposes.
	Right: Standard deviations for $\overline{p}_k$ and $\overline{x}_k$ in Eqs.~(\ref{sigp}) and~(\ref{sigx}).}
        \label{momsnonhom}
\end{figure}

In order to put in evidence the very large value of the ergodic time $t_e$ we implemented a molecular dynamics simulation of an open
$N$-particle system (no spatial boundary conditions) with Hamiltonian in Eq.~(\ref{ham1dgrav}) using and event-driven algorithm~\cite{allen}. The dynamics
between two successive particle crossings is integrable, and can be computed up to
machine precision. Collisions are then implemented straightforwardly by updating the force on the particles after each crossing. Due to
very high local densities at the core of the spatial distribution, a high numeric precision is required and we used
quadruple precision in order to avoid missing
any collision due to round-off errors (which indeed occur for double precision). The initial state is a waterbag state defined by
\begin{equation}
	f(x,p;0)=\left\{
		\begin{array}{l}
			1/p_0r_0,\hspace{3mm}{\rm if}\,\, -x_0<x<x_0\,{\rm and}\,-p_0<p<p_0,\\
			0,\,\,\,\,{\rm otherwise},
		\end{array}
		\right.
	\label{waterbagdef}
\end{equation}
with $x_0$ and $p_0$ given constants. To measure the distance to the Gaussian distribution we use the reduced moments:
\begin{equation}
	\mu_k\equiv\frac{\langle p^k\rangle}{{\langle p^2 \rangle}^{k/2}}.
	\label{redmomdef}
\end{equation}
The reduced moment of order 4 is called the Kurtosis of the distribution, and for any Gaussian distribution we have that $\mu_4=3$ and $\mu_6=15$.
The left panel of Fig.~\ref{momsnonhom} shows the time evolution of $\mu_4$ and $\mu_6$ for the system,
with $x_0=10.0$ and $p_0=0.5$ for the initial condition. In this case the relaxation time to equilibrium is of the order of $t_r\approx10^6$.
The right panel of Fig.~\ref{momsnonhom} shows that the condition for ergodicity stated in Eq.~(\ref{defergosig}) is satisfied
for a value of time of the order of magnitude of the relaxation time for equilibrium $t_e\approx t_r$.

\begin{figure}[ht]
	\begin{center}
	\scalebox{0.3}{\includegraphics{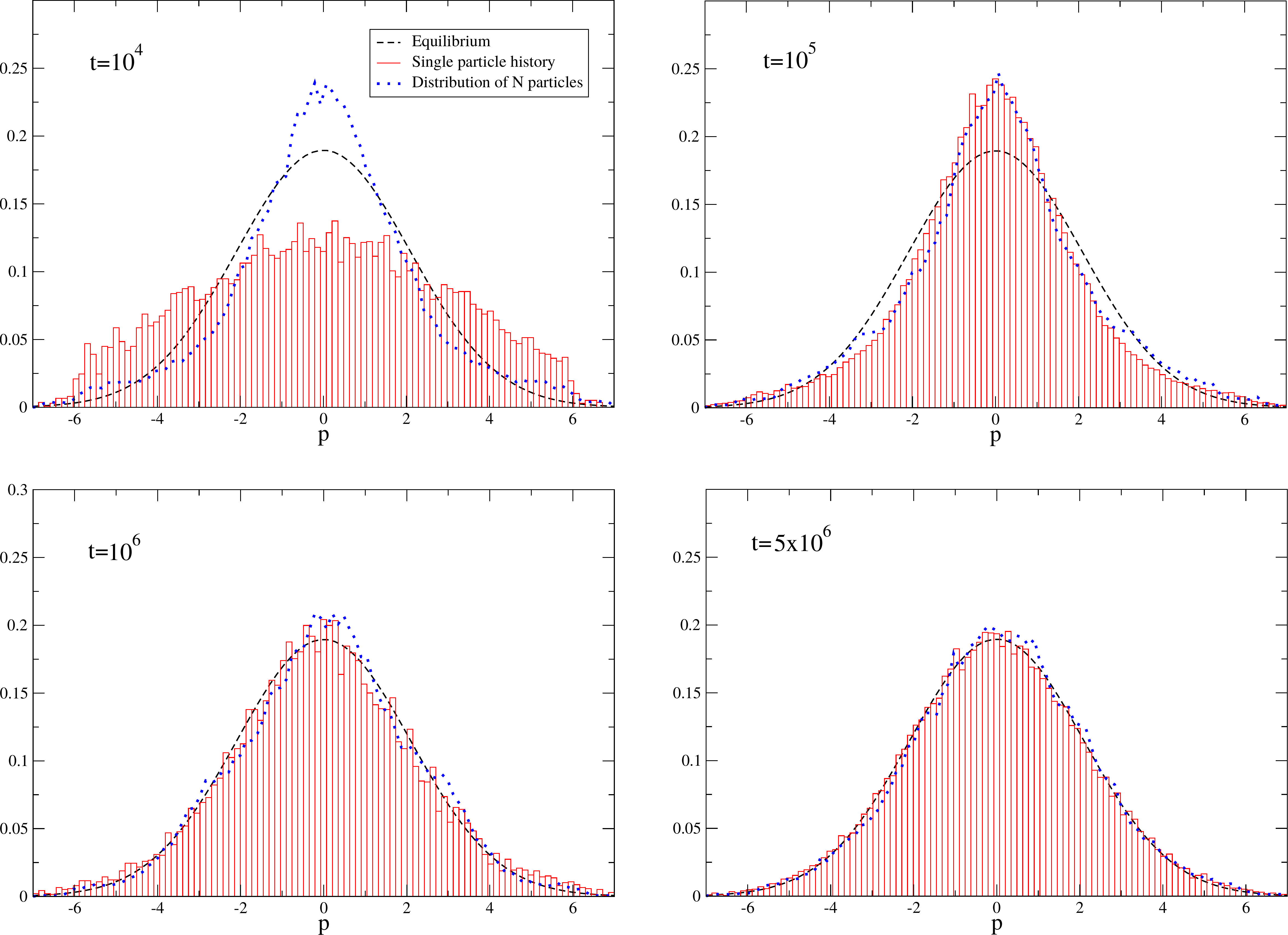}}
	\end{center}
%        \scalebox{0.3}{\includegraphics{histogram0.pdf}}
%	\scalebox{0.3}{\includegraphics{histogram1.pdf}}
%	\scalebox{0.3}{\includegraphics{histogram2.pdf}}
%	\scalebox{0.3}{\includegraphics{histogram3.pdf}}
	\caption{Distributions $\phi(p;t)$ (dotted line) and $g(p;t)$ (histogram) for the same simulation as in Fig.~\ref{momshom}
	and for few values of $t$. The dashed line is the equilibrium Gaussian with $\beta=0.225$. This value of $\beta$
	was obtained by averaging the kinetic energy 	for a time window of size $\delta t=10\,000$ at the end of the simulation.
	The precision for the histogram for $p_k(t)$ was increased by collecting the values of the momenta of all particle from
	time $t-100$ up to $t$, justified by an expected negligible change in the statistical distribution for a relatively
	short period of time.}
        \label{distmom}
\end{figure}

\begin{figure}[ht]
	\begin{center}
	\scalebox{0.3}{\includegraphics{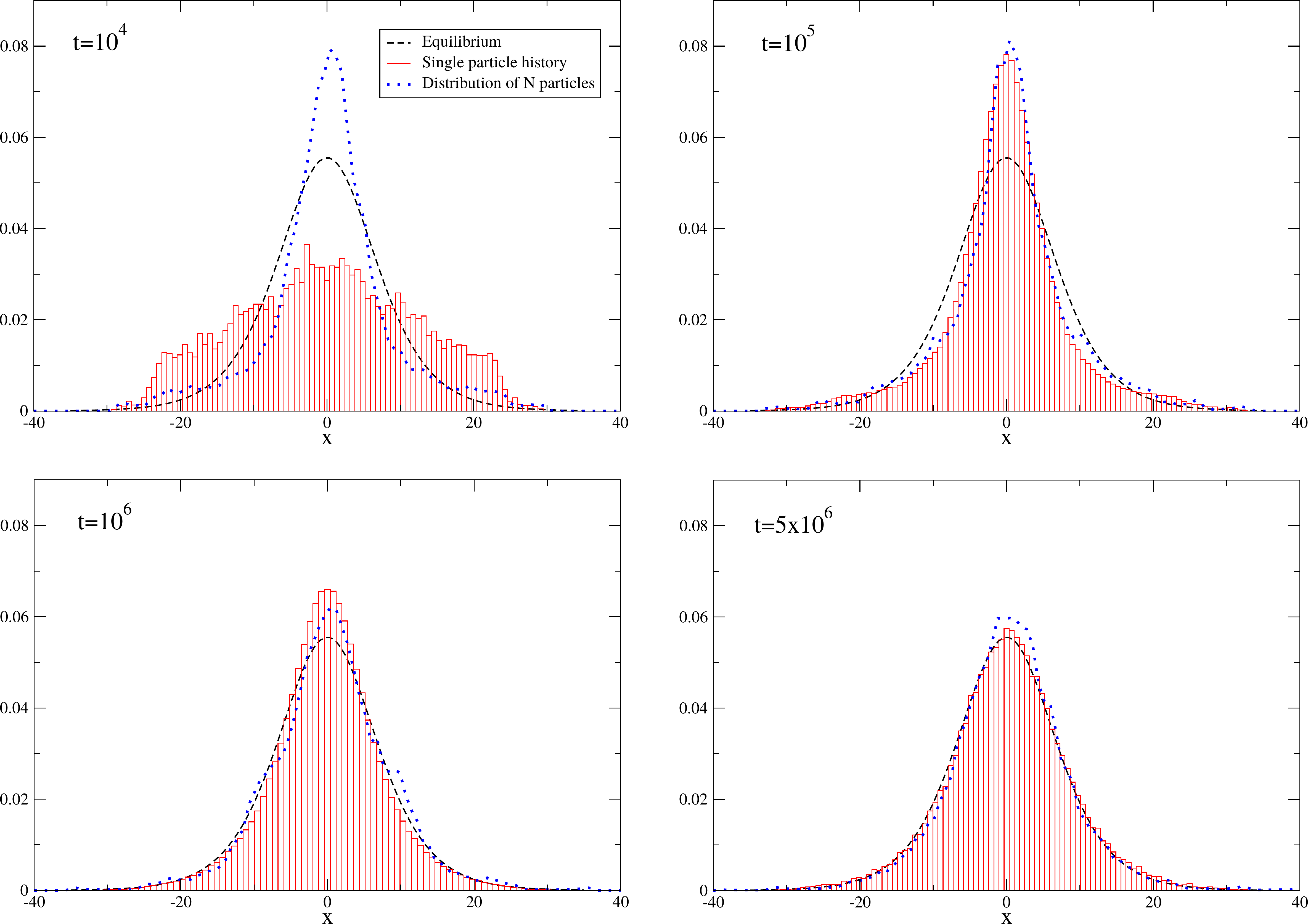}}
	\end{center}
%        \scalebox{0.3}{\includegraphics{histogram_x0.pdf}}
%        \scalebox{0.3}{\includegraphics{histogram_x1.pdf}}
%        \scalebox{0.3}{\includegraphics{histogram_x2.pdf}}
%        \scalebox{0.3}{\includegraphics{histogram_x3.pdf}}
	\caption{Same as Fig. \ref{distmom} but for $h_t(x)$ and $\rho(x;t)$. The dashed line
	is the spatial distribution function at equilibrium $\rho(x)=C {\rm sech}(x/\Lambda)$, $\Lambda=4e/3$ and $e$
	the mean-field energy per-particle~\cite{rybicki}.}
        \label{distpos}
\end{figure}

In order to discuss the physical meaning of ergodicity for a long range interacting system,
we define the one-particle momentum and position probability densities $\phi(p;t)$ and $\rho(x;t)$ at a given time $t$
as the probability density for the given value of $p$ and $x$, respectively.
We also define the density distribution for the values of $p$ and $x$ for a fixed particle,
say the $k$-th particle, along its history, up to time $t$,
denoted by $g(p;t)$ and $h(x;t)$ respectively.
Then, in the present case, ergodicity is equivalent to the relations
\begin{equation}
	\phi(p;t)=g(p;t),
	\label{phig}
\end{equation}
and
\begin{equation}
	\rho(x;t)=h(x;t),
	\label{rhoh}
\end{equation}
for $t\gtrapprox  t_r\approx t_e$.
Figures~\ref{distmom} and~\ref{distpos} show these distributions for a few values of time, and also the spatial distribution function
at equilibrium given by $\rho(x)=C {\rm sech}(x/\Lambda)$, with $\Lambda=4e/3$ and $e$ the mean-field energy per-particle~\cite{rybicki},
and the momentum Gaussian distribution at equilibrium.
It is evident that the time and ensemble distributions become very close as $t$ approaches $t_r$. So the momentum $\phi$ and $g$, and spatial $\rho$ and $h$,
distribution functions satisfy Eqs.~(\ref{phig}) and~(\ref{rhoh}) and are also equal to the equilibrium distribution for a time of the order of magnitude
of the relaxation time to equilibrium, as it was also observed for other long range interacting systems~\cite{ergo1,ergo2,ergo3}.

\subsection{Homogeneous state}

\begin{figure}[ht]
	\begin{center}
	\scalebox{0.3}{\includegraphics{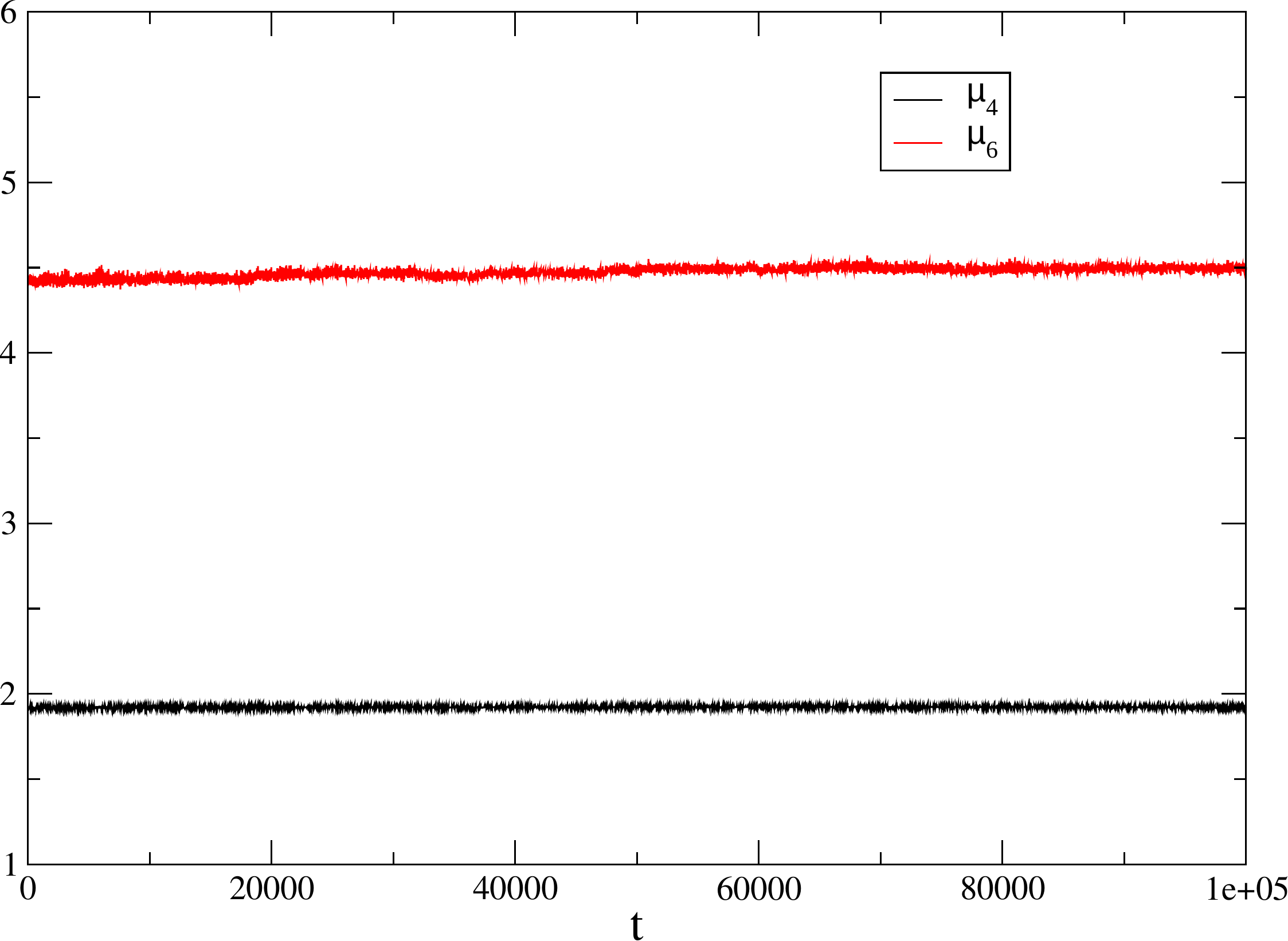}}
	\end{center}
%        \scalebox{0.3}{\includegraphics{moments_hom.pdf}}
        \caption{Reduced moments of $p$ for a homogeneous state with a waterbag initial condition with $x_0=1.0$ and $p_0=3.0$.}
        \label{momshom}
\end{figure}

\begin{figure}
	\begin{center}
	\scalebox{0.3}{\includegraphics{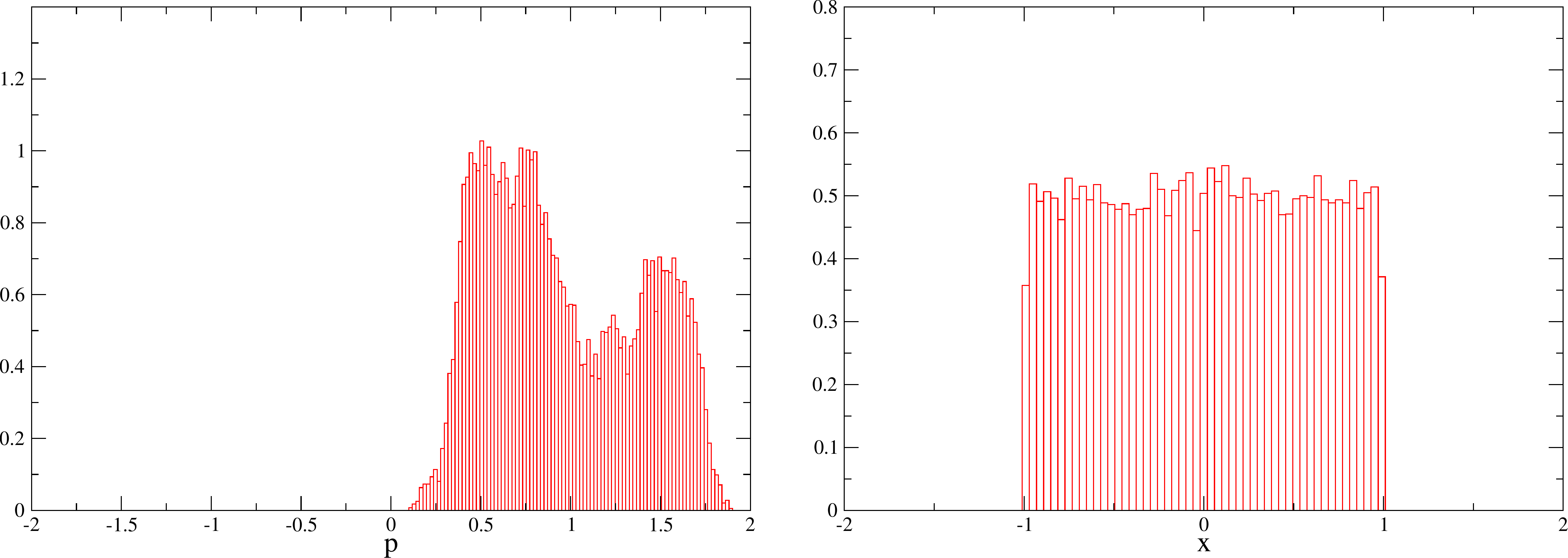}}
	\end{center}
%        \scalebox{0.3}{\includegraphics{histogram_p_hom.pdf}}
%        \scalebox{0.3}{\includegraphics{histogram_x_hom.pdf}}
	\caption{Left: Distribution function $g_t(p)$ for $t=10^5$ for the same simulation as in Fig.~\ref{distspxhom}.
	Right: distribution function $\rho_t(x)$ for the same simulation.}
	\label{distspxhom}
\end{figure}

We now turn to the case of a homogeneous state. Periodic boundary conditions can be implemented using an Ewald sum with a unit
cell $x\in[-L,L]$ such that the force on each particle, due to the particles in the unit cell and the infinite number of images, is determined
by a direct sum over replicas~\cite{hernquist}. For the one-dimensional self-gravitating system a closed analytical form was obtained
by Miller and Rouet~\cite{miller2} as an additional potential representing all replicas, and given by:
\begin{equation}
        V_{\rm Ewald}=-\frac{1}{N}\sum_{i=1}^N\frac{(x-x_i)^2}{2L}.
        \label{ewaldsumpot}
\end{equation}
The full effective Hamiltonian with periodic boundary conditions is then
\begin{equation}
        H=\sum_{i=1}^N\frac{{\bf p}_i^2}{2m}+V({\bf x}),
\label{genlongham2}
\end{equation}
with ${\bf x}\equiv(x_1,\ldots,x_N)$ and
\begin{equation}
        V({\bf x})=\frac{1}{N}\sum_{i<j=1}^N\left|x_i-x_j\right|-\frac{1}{N}\sum_{i=1}^N\frac{(x-x_i)^2}{2L}.
\label{poteff}
\end{equation}
The resulting equations of motion are then integrated using a fourth order symplectic integrator~\cite{yoshida,eu2}.
The  reduced moments $\mu_4$ and $\mu_6$ as a function of time, up to $t=10^5$, are shown in Fig.~\ref{momshom},
for an initial waterbag state with $x_0=L=1$ and $p_0=3$. The system remains in a homogeneous state for the whole simulation time.
We observe that the time evolution is
extremely slow if compared to the non-homogeneous case, with only a very small variation in $\mu_6$ visible in the graphic.
Figure~\ref{distspxhom} shows the distribution functions
$g(p;t)$ and $g(x;t)$ at the final time, also clearly at variance to what is observed for the non-homogeneous cases. Although the spatial distribution
$g(x;t)$ is roughly uniform, as expected since particles can cross each other and the mean-field force is very small, the momentum
distribution $g(p;t)$ is not even symmetrical, as it is the case for the non-homogeneous systems at all time values, except for a very
short initial time. We conclude that the time for ergodicity, if finite, is certainly many orders of magnitude greater that
for a non-homogeneous state.
We will shed some light and explain the physical origin of this difference, and of the peculiar dynamics of the homogeneous state,
in the next section by discussing the kinetic theory for a homogeneous state.

\section{Kinetic equation for a homogeneous state}
\label{sec3}

The statistical dynamics of a system of many particle systems can be studied by determining a kinetic equation describing the
time evolution of the one-particle distribution function.
We first define the $N$-particle distribution function $f_N(x_1,p_1,\ldots,x_n,p_n;t)$ as the probability density in the
$N$-particle phase space, which satisfies the Liouville equation. 
An usual starting point to determine a kinetic equation is the BBGKY hierarchy for the reduced
distribution functions~\cite{balescu,liboff}:
\begin{eqnarray}
	\frac{\partial}{\partial t}f_s(1,\ldots,s;t) & = & -\sum_{k=1}^s p_k\frac{\partial}{\partial x_k}f_s(1,\ldots,s;t)
+\frac{1}{2}\sum_{\substack{k,l=1\\(k\neq l)}}^sV_{kl}^\prime\:\partial_{kl}f_s(1,\ldots,s;t)
\nonumber\\
	& & +(N-s)\sum_{k=1}^s\int\dd(s+1)\:V_{k,(s+1)}^\prime\:\frac{\partial}{\partial p_k}f_{s+1}(1,\ldots,s+1;t),
\label{bbgky}
\end{eqnarray}
where $V_{jk}\equiv V(x_j-x_k)$, $\partial_{kl}\equiv p_k\partial/\partial x_k-p_l\partial/\partial x_l$, $1\equiv x_1,p_1$, $2\equiv x_2,p_2$, \ldots,
$\dd1\equiv \dd x_1\dd p_1$, $\dd2\equiv \dd x_2\dd p_2$, \ldots, and so on. The $s$-particle distribution function is defined by:
\begin{equation}
        f_s(1,\ldots,s;t)\equiv\int\dd1\cdots\dd (s+1)\:f_N(1,\ldots,N;t).
\label{reddist}
\end{equation}
The case with $s=1$ leads to the prototypical kinetic equation:
\begin{equation}
        \left[\frac{\partial}{\partial t}+p_1\frac{\partial}{\partial x_1}\right]f_1(1;t)=
        (N-1)\frac{\partial}{\partial p_1}\int\dd2\:V_{12}^\prime\:f_2(1,2;t).
        \label{protkineq}
\end{equation}
In order to obtain a close-form expression for the kinetic equation one must determine
an expression for the two-particle distribution $f_2$ in terms of $f_1$. For uncorrelated particles
we have $f_2(1,2;t)=f_1(1;t)f_1(2;t)$ and Eq.~(\ref{protkineq}) then results in the Vlasov equation~(\ref{vlasoveq}).

For non-correlated particles we perform the cluster expansion~\cite{balescu}:
\begin{eqnarray}
	f_2(1,2;t) & = & f_1(1;t)f_1(2;t)+g_2(1,2;t),\nonumber\\
	f_3(1,2,3;t) & = & f_1(1;t)f_1(2;t)f_1(3;t)+f_1(1;t)g_2(2,3;t)+f_1(2;t)g_2(1,3;t)\nonumber\\
	 & & +f_1(3;t)g_2(1,2;t)+g_3(1,2,3;t),
	\label{clusterexp}
\end{eqnarray}
and so on, where $g_s$ is the $s$-particle correlation function. By plugging Eq.~(\ref{clusterexp}) into Eq.~(\ref{bbgky})
for $s=1$ we have
\begin{equation}
	\frac{\partial }{\partial t}f_1(1;t)=N\int \dd2\:V^\prime_{12}\partial_{12}
\left[f_1(1;t)f_1(2;t)+g_2(1,2;t)\right].
\label{protkin}
\end{equation}
From Eqs.~(\ref{bbgky}), (\ref{clusterexp}) and (\ref{protkin}) we obtain the following equation for the two-particle correlation function~\cite{balescu}:
\begin{eqnarray}
	\left(\frac{\partial }{\partial t}+p_1\frac{\partial}{\partial x_1}+p_2\frac{\partial}{\partial x_2}\right)g_2(1,2,t) & = &
	V^\prime_{12}\partial_{12}f_1(1;t)f_1(2;t)+V^\prime_{12}\partial_{12}g_2(1,2;t)\nonumber\\
	& & \hspace{-20mm}+N\int\dd3\left[V^\prime_{13}\partial_{13}f_1(1;t)g_2(2,3;t)+V^\prime_{23}\partial_{23}f_1(2;t)g_2(1,3;t)\right.\nonumber\\
	& & \hspace{-20mm}\left.+\left(V^\prime_{13}\partial_{13}+V^\prime_{23}\partial_{23}\right)\left\{f_1(3;t)g_2(1,2;t)+g_3(1,2,3;t)\right\}\right].
\label{eqg2}
\end{eqnarray}
Then one determines a solution for $g_2$ in terms of $f_1$, and also a solution for $g_3$ if it is
not negligible (see~\cite{scaling} and~\cite{fouvry} for systems where three-particle correlations are important),
and use the result in Eq.~(\ref{protkineq})

\begin{figure}
	\begin{center}
	\scalebox{0.3}{\includegraphics{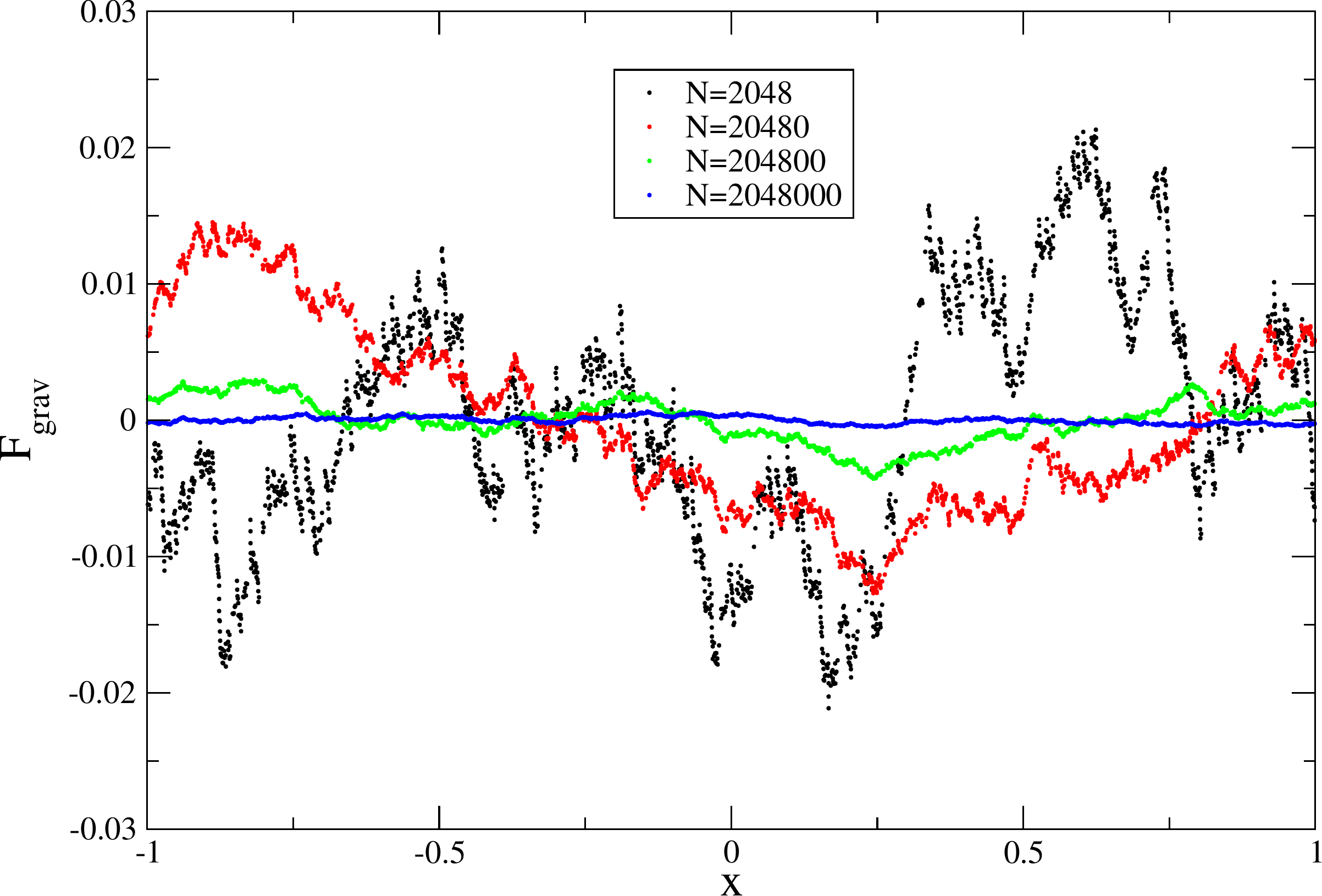}}
	\end{center}
%        \scalebox{0.3}{\includegraphics{force_homogeneous2.pdf}}
	\caption{Force as a function of position for the Hamiltonian in Eq.~(\ref{genlongham2}) for the one-dimensional self-gravitating
	system with an additional potential from the Ewald sum, for different number of particles but same density $n=N/L$.
        The position was rescaled to the interval $[-1,1]$ for comparison purposes. We unit cell for $N=2048$ is given by $L=20$,
	and is obtained accordingly for other values of $N$ in order to keep the spatial density constant.}
        \label{forcehom}
\end{figure}

Now we turn to the one-dimensional self-gravitating system with Hamiltonian in Eq.~(\ref{ham1dgrav}) in a homogeneous state.
The derivative of the total potential $V$ in Eq.~(\ref{poteff}) appears in Eq.~(\ref{eqg2}) and
we must consequently account for the singularity of its derivative at zero inter-particle distance.
We consider the following relabeling of particle indices: at the moment two particles (sheets) cross each other, we interchange their labels.
In this way, at each collision (at zero distance) particles simple exchange their momenta and the force is constant in time.
If the particles are initially labeled such that $x_i<x_j$ if $i<j$, the ordering in position is preserved. Then the force on particle $i$ due
to particle $j$ can be written as $F=F_{\rm grav}+F_{\rm HC}$ where $F_{\rm grav}=-V^\prime(x_i-x_j)$, with $V$ given in Eq.~(\ref{poteff}), and
$F_{\rm HC}$ stands for the hard-core force that swaps particle momenta when they collide at zero distance.
The contribution of $F_{\rm grav}$ to Eq.~(\ref{eqg2}) vanishes in the limit $L\rightarrow\infty$ as the gravitational force in a homogeneous state
vanishes. To illustrate this fact, Figure~\ref{forcehom} shows the force $F_{\rm grav}$ due to both the self-gravitating potential
and the Ewald sum, for a few values of the number of particles $N$ but for keeping the density $n=N/L$ constant.
We note that increasing $N$ in this way is not equivalent to consider the thermodynamic limit that would correspond to take $N\rightarrow\infty$
but keeping $L$ constant. We observe that as the size $L=N/n$ of the unit cell increases
$F_{\rm grav}$ approaches zero. As a consequence, only contributions from hard-core collisions are retained in Eq.~(\ref{eqg2}).
This result in fact proves the validity of the Jeans Swindle for the model considered here, i.~e.\ that the contribution of the background interaction
to the infinite homogeneous contribution vanish, and one must consider only the effects of local fluctuations in density~\cite{kiessling,falco}.
These fluctuations vanish as the size of the unit cell goes to infinity.

The same reasoning can be used in an analogous way for the BBGKY hierarchy, which then take exactly the same form
as the hierarchy obtained for a system of particles with a hard-core potential at zero distance as only interaction.
For such a system in an homogeneous state, the one-particle distribution function $f_1(p;t)$ is strictly constant
in time as the interaction only swaps the momenta of two particles at each collision,
and three-particles processes are nonexistent (the probability that three particles collide at the same time at the same point is zero).
For the same initial condition, the BBGKY hierarchy being identical for both systems,
the time evolution for the reduced distribution functions must be the same, and therefore the distribution $f_1(p;t)$ for
a homogeneous one-dimensional self-gravitating system is constant in time. Small deviations from this are expected to occur in numerical simulations
due to spurious non-physical effects resulting from a finite value of $L$, that result in small fluctuations of the value of the force around zero.

\section{Concluding Remarks}
\label{sec4}

We showed that the sheets model describing a one-dimensional self-gravitating system has profoundly different dynamic properties
weather it is in a homogeneous or a non-homogeneous state.
In the former case we showed that by considering a proper limit in the periodic boundary conditions the one-particle evolution function
does not evolve in time, as its kinetic equation is essentially a Boltzmann-like equation.
For the non-homogeneous state,
the system has a slow dynamics to equilibrium, with a relaxation time much greater than other long range interacting systems if one uses
the violent relaxation time for comparison. The non-homogeneous system is ergodic but only after a time of the order of the relaxation time
to equilibrium, as also observed for other long range interacting systems, but it is non-ergodic in a homogeneous state, as illustrated by
simulations presented here.

A possible way to shed some light on the slow dynamics of this system in a non-homogeneous states is to
obtain a kinetic equation, which for the present model is a challenging task as it requires the
determination of action-angle variables for the mean-field description of the system~\cite{CHAVANIS20123680,CHAVANIS2007469},
and has been possible only for very special cases (see~\cite{benetti} and references therein).
This is the subject of ongoing research.

\section*{Acknowledgments}

LFS was financed by CNPq (Brazil). TMRF was partially financed by CNPq under grant no.\ 305842/2017-0.

\end{document}